%% file: paper.tex
\documentclass[conference]{IEEEtran}
\IEEEoverridecommandlockouts
\usepackage{cite}
\usepackage{amsmath,amssymb,amsfonts}
\usepackage{algorithmic}
\usepackage{graphicx}
\usepackage{textcomp}
\usepackage{multirow}
\usepackage{xcolor}
\usepackage{booktabs}
\usepackage{eso-pic}
\usepackage{threeparttable}
\usepackage[multiple]{footmisc}
\usepackage{todonotes}
\usepackage{balance}
\def\BibTeX{{\rm B\kern-.05em{\sc i\kern-.025em b}\kern-.08em
    T\kern-.1667em\lower.7ex\hbox{E}\kern-.125emX}}

\usepackage{tikz}
\usepackage[hyperfootnotes=false]{hyperref}

\newcommand\copyrighttext{%
  \footnotesize \textcopyright 2023 IEEE. Personal use of this material is permitted.
  Permission from IEEE must be obtained for all other uses, in any current or future
  media, including reprinting/republishing this material for advertising or promotional
  purposes, creating new collective works, for resale or redistribution to servers or
  lists, or reuse of any copyrighted component of this work in other works.
  DOI: \href{https://doi.org/10.1109/BigData59044.2023.10386195}{https://doi.org/10.1109/BigData59044.2023.10386195}}
\newcommand\copyrightnotice{%
\AddToShipoutPicture*{%
\put(45,30){%
\centering
\fbox{\parbox{\dimexpr\textwidth-\fboxsep-\fboxrule\relax}{\copyrighttext}}
}}}

\DeclareMathAlphabet{\altmathcal}{OMS}{cmsy}{m}{n}
    
\begin{document}

\title{Towards a Peer-to-Peer Data Distribution Layer for Efficient and Collaborative Resource Optimization of Distributed Dataflow Applications}

\author{
\IEEEauthorblockN{
Dominik Scheinert\IEEEauthorrefmark{1},
Soeren Becker\IEEEauthorrefmark{1},
Jonathan Will\IEEEauthorrefmark{1},
Luis Englaender\IEEEauthorrefmark{1},
and Lauritz Thamsen\IEEEauthorrefmark{2}
}
\IEEEauthorblockA{
\IEEEauthorrefmark{1}
Technische Universit{\"a}t Berlin, Germany, \{firstname.lastname\}@tu-berlin.de}
\IEEEauthorblockA{
\IEEEauthorrefmark{2}
University of Glasgow, United Kingdom, lauritz.thamsen@glasgow.ac.uk}}

\maketitle
\copyrightnotice

\begin{abstract}
\input{sections/0_abstract}
\end{abstract}


\input{sections/1_introduction}
\input{sections/2_system_idea}
\input{sections/3_data_distribution_layer}
\input{sections/4_evaluation}

\input{sections/5_related_work}
\input{sections/6_conclusion}

\section*{Acknowledgments}
This work has been supported through grants by the Deutsche Forschungsgemeinschaft (DFG, German Research Foundation) as C5 (grant 506529034) and as FONDA (Project 414984028, SFB 1404).

\bibliographystyle{IEEEtran}
\bibliography{bib}
\balance

\end{document}

%% file: sections/0_abstract.tex
Performance modeling can help to improve the resource efficiency of clusters and distributed dataflow applications, yet the available modeling data is often limited.
Collaborative approaches to performance modeling, characterized by the sharing of performance data or models, have been shown to improve resource efficiency, but there has been little focus on actual data sharing strategies and implementation in production environments.
This missing building block holds back the realization of proposed collaborative solutions.

In this paper, we envision, design, and evaluate a peer-to-peer performance data sharing approach for collaborative performance modeling of distributed dataflow applications.
Our proposed data distribution layer enables access to performance data in a decentralized manner, thereby facilitating collaborative modeling approaches and allowing for improved prediction capabilities and hence increased resource efficiency.
In our evaluation, we assess our approach with regard to deployment, data replication, and data validation, through experiments with a prototype implementation and simulation, demonstrating feasibility and allowing discussion of potential limitations and next steps.

%% file: sections/1_introduction.tex
\section{Introduction}
\label{sec:introduction}

Apache Spark~\cite{ZahariaCFSS10} and Apache Flink~\cite{CarboneKEMHT15} are two prominent examples of distributed dataflow frameworks that enable the handling of large datasets, originating from diverse sources such as web logs, sensor data, e-commerce transactions, or clickstream data, using clusters of commodity hardware by implementing failure handling and parallel execution.
The optimized configuration of the underlying cluster resources, however, is a non-trivial task~\cite{RzadcaFSZBKNSWH20,AlDhuraibiPDM18,RajanKCK16} that is commonly approached with performance modeling techniques~\cite{KirchoffXMR19,AlSayehS19,ChenLLWZ21silhouette,bilal2020finding,MendesCRG20,CasimiroD0RZG20,ScheinertTZWAWK21}.
Nevertheless, they usually require a sufficient amount of previously recorded training data~\cite{BaderWBLLDVK22} or dedicated profiling~\cite{BeckerSSK22} to accurately model the configuration-dependent application behavior, rendering the solutions impractical in many situations.

At the same time, many distributed dataflow applications share key characteristics, e.g., in terms of employed algorithm implementations, distributions in datasets, or infrastructures, which presents an opportunity for collaborative approaches to performance modeling.
Recent related~\cite{FekryCPR20,FekryCPRH20} and previous own works~\cite{ScheinertABWWT21,WillTSBK21} have presented ideas that are generally suitable for a collaborative scenario. Still, their main focus is on the respective modeling strategies, not so much on the architecture, design decisions, and implications surrounding the actual data sharing.
Sketched solutions furthermore advertise a centralized data sharing approach, which comes with potential drawbacks. The centralized data sharing approach might not be able to cope with the increasing data scale needed to generate robust performance models. Additionally, providing and maintaining a centralized storage system for large-scale performance data sets involves significant costs, i.e. for storage, bandwidth, or personnel of the central institution. This limitation of current collaborative approaches could be hindering real integration and usage.

In order to address these challenges, we believe that a peer-to-peer (P2P) data distribution layer provides a compelling alternative since it enables efficient network traffic distribution among all participating peers~\cite{DBLP:conf/ipccc/Becker0K21}, fosters data sovereignty~\cite{DanielT22}, and reduces the risk of a single point of failure. Additionally, since data can be replicated across nodes in the P2P network in an ad-hoc manner, reliability is improved without the need for expensive central server infrastructure~\cite{NaikK20}. 
In this paper, we present our vision of a P2P data distribution layer that enables a scalable and global exchange of performance datasets without geographical limitations.
This facilitates collaboration while eliminating the need for a single, controlling authority or storage infrastructure.
\vspace{2mm}

\hspace{-4mm}\emph{Contributions.} The contributions of this paper are:

\begin{itemize}
    \item We outline collaboration as an important preliminary for many resource optimization use cases and derive requirements for a decentralized data distribution layer.
    \item We propose a system design for a decentralized data distribution layer based on peer-to-peer networks.
    \item We conduct an empirical evaluation in which we assess our architecture using a prototype implementation and a simulation and discuss preliminary findings.
\end{itemize}

\vspace{2mm}

\hspace{-4mm}\emph{Outline}. The remainder of the paper is structured as follows.
\autoref{sec:system_idea} presents the idea and proposes a decentralized system for collaboratively optimizing resource configurations of distributed dataflows. 
\autoref{sec:data_distribution_layer} further explains the necessary data distribution layer and its role as well as integration in the overall collaborative performance modeling system.
\autoref{sec:evaluation} presents the preliminary results of our conducted experiments and a discussion of general requirements of our approach.
\autoref{sec:related_work} discusses the related work.
\autoref{sec:conclusion} concludes the paper.

%% file: sections/2_system_idea.tex
\section{System Idea}
\label{sec:system_idea}

\begin{figure}[h!]
    \centering
    \includegraphics[width=.9\columnwidth]{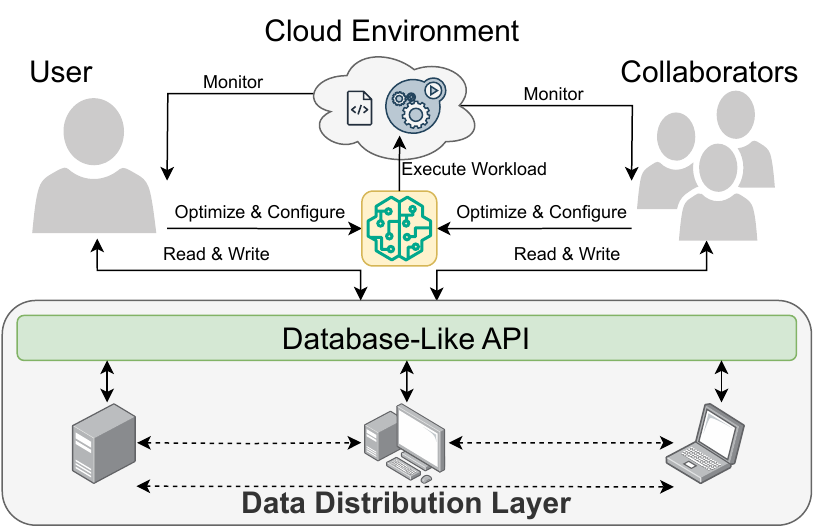}
    \caption{Overview of the envisioned system.
    From the user's perspective, sharing and collecting data is abstracted away and takes place under the hood, so that the attention is directed toward performance modeling of dataflows.}
    \label{fig:system_overview}
\end{figure}

This section elaborates on our envisioned system for collaborative performance modeling of distributed dataflow applications through data sharing in a decentralized and ad-hoc manner. 
An overview of the system is provided in~\autoref{fig:system_overview}.

\subsection{Decentralized Performance Data Sharing}

Resource optimization for distributed dataflow applications via performance modeling is a challenging endeavor as the required amounts of training data are oftentimes difficult to obtain.
This can be overcome with collaborative approaches that build upon the sharing of performance data and/or models.
Yet, existing approaches assume a centralized storage solution, which introduces additional challenges.
On the one hand, the scalability of centralized solutions must be questioned in light of increasing data amounts, and it is also often not in line with the needs of collaborators, who might be interested in particular subsets of data only.
On the other hand, apart from concerns regarding fault tolerance, centralized solutions also induce substantial costs and responsibilities for a single party.
Lastly, as collaborators might also want to withhold certain sensitive data, it would be necessary to operate two data sources either way.
Hence, for such approaches, it is interesting to explore the feasibility of peer-to-peer sharing strategies to enable decentralized and collaborative performance data access and provisioning. 

\subsection{Envisioned System}

We envision a decentralized data distribution layer that loosely connects various participating peers.
Peers can sporadically join or leave the network, but share a joint intrinsic motivation in gaining access to valuable workload performance data for improved performance modeling.
At its core, the idea is that the performance modeling approach utilized by a particular party interacts with an API that abstracts away aspects such as data retrieval, replication, or access to differing data sources.
While the API offers simplified means of loading and saving performance data, the remaining parts of the data distribution layer take care of the distribution, replication, and validation of individual performance data points. 
Crucially, using the data distribution layer, users retain control over when and what data is shared through configuration options.

%% file: sections/3_data_distribution_layer.tex
\section{Data Distribution Layer}
\label{sec:data_distribution_layer}

\begin{figure}[h!]
    \centering
    \includegraphics[width=\columnwidth]{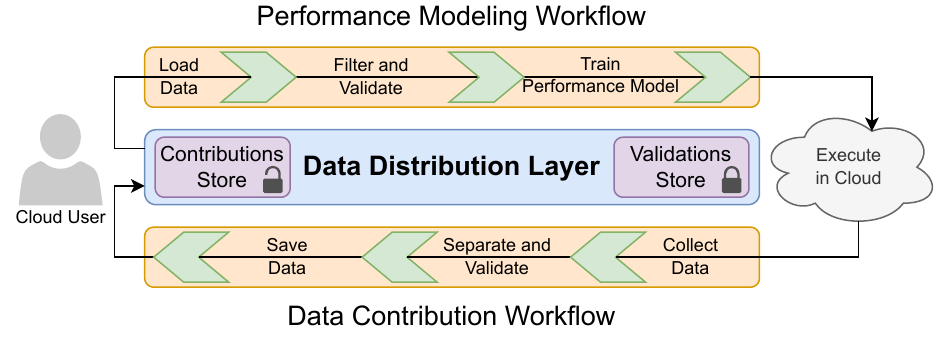}
    \caption{The proposed workflows and actions. The data distribution layer represents a key component and facilitates data management and access for downstream tasks such as performance modeling for resource allocations.}
    \label{fig:approach_overview}
\end{figure}

This section presents our approach to a decentralized data distribution layer for the sharing of performance data and collaboratively optimizing resource configurations of distributed dataflow applications. 
It encompasses multiple workflows and steps, which are sketched in~\autoref{fig:approach_overview}.

\subsection{Background on the Interplanetary File System}
As one of the primary foundational elements of our approach, we have adopted the InterPlanetary File System (IPFS)~\cite{benetIPFSContentAddressed2014}.
IPFS is a P2P version-controlled distributed filesystem that leverages the Kademlia Distributed Hash Table (DHT)~\cite{maymounkovKademliaPeertoPeerInformation2002} to facilitate the discovery of network addresses pertaining to peer nodes and the IPFS objects hosted by said peers. Objects can be uniquely identified through a cryptographic representation of their content, denoted as a content identifier (CID), that permits other peers to efficiently retrieve the associated object in a location-agnostic way. Any peer hosting a given object seamlessly shares it with fellow peers upon request~\cite{DanielT22}. In addition, IPFS also provides IPFS-Log, an append-only log that can be used to model a mutable, shared state between peers in P2P applications, as it is an operation-based conflict-free replicated data type (CRDT). In general, due to the properties of IPFS, e.g. tamper-resistance, deduplication, and decentralization, it is frequently combined with blockchains~\cite{KumarBA21} and utilized as one of the main building blocks of highly scalable decentralized systems~\cite{ChenLLZ17}.

\subsection{Data Organization and Distribution}

To foster the integration of our data distribution layer with existing approaches to performance modeling that utilize database-backed solutions, we require a comparable API with database-like operations, backed by IPFS, and the possibility to seamlessly access different types of data: local and private data, as well as global and shared data.
Assuming that each peer runs its own instance of IPFS for data storage, we can support multiple use cases: On the one hand, data can be stored that is not intended for sharing. This can be sensitive and detailed monitoring information, which might be of interest to the respective user (and hence locally persisted) but not expected to be shared.
On the other hand, publicly shared data in the network can be pinned and as a consequence replicated to a local IPFS node.
This enables flexibility and autonomy. 

For publicly shared, meaningful performance data, which we will refer to as \emph{contributions} in the following, we employ a data structure called \emph{contributions store} based on IPFS-Log.
The contributions store is ideally replicated among all peers and contains CIDs of actual performance data. 
These CIDs can then be used to retrieve the actual data from the P2P network, keeping the contribution store compact and easy to navigate.
In principle, the data format of the contributions store could also be extended with additional attributes, e.g., in order to filter CIDs by cloud platforms the performance data was gathered on.
As for private data, it can directly be stored on the local IPFS node, which enables relatively strong privacy as long as the corresponding CID is not shared with a third party.
For additional security, a middleware can be employed that denies external CID requests for particular CIDs.

\subsection{Data Validation and Integrity}

So far we discussed the organization and distribution of data, yet validation and integrity are still to be tackled.
Fortunately, data integrity is a built-in feature of IPFS due to the concepts behind content-addressing which enable tamper-resistance while not enforcing any specific data format.

Performance data validation can be addressed on multiple levels: A straightforward step is the implementation of access control, i.e., the requirement of a passphrase for joining through the IPFS bootstrapping node.
Still, corrupted or inconclusive data could be contributed without malicious intent. 
In order to account for this, validation routines can be employed, both before insertion as well as after replication. 
These validation routines are in turn ideally composed of actions that validate data quality as well as the benefit for performance modeling~\cite{dataqualityML22,WillABST21}. 
To standardize these processes, the required code for the validation pipelines can also be stored in IPFS, and shared among peers.
To reduce the overhead of validation on an individual level, several strategies qualify: One option would be to e.g. organize the validation results of particular CIDs based on a blockchain.
In this initial work, we follow a more lightweight and opportunistic approach: A user requests the individual validation results of other peers in the network and consolidates them -- in case of an inconclusive vote or undesired outcome, the performance data of interest is validated independently, otherwise the decision of the network is used.
Generally, each user maintains a local data structure in IPFS with validation results for particular CIDs, called \emph{validations store}, which can be consulted if needed or used to share validation data with other peers upon request.

\subsection{Performance Modeling Workflow}

Whenever a new performance model shall be trained for the purpose of resource configuration optimization of distributed dataflows, the contributions store is consulted and the required performance data is retrieved by their CIDs, either from a local IPFS node or remote IPFS nodes.
The contributions are optionally pre-filtered according to further criteria, or based on their data validity which is either decided from knowledge in the P2P network or by a user's own validation steps.
Previously only remotely available data could then be marked as qualifying for IPFS pinning, to reduce latencies for future training and to share it with other peers.
The gathered data contributions can additionally be joined with performance data which is only locally available, and eventually used for training and employment of a performance model.
The outlined internals shall be masqueraded and automated for a user, to reduce complexity and foster usage of our approach.

\subsection{Data Contribution Workflow}

Contribution is a crucial aspect of any sharing-based project.
As for our envisioned system, contributing shall be an automated~\cite{ScheinertABWWT21}, yet configurable step executed after each run of a distributed dataflow application.
This achieves long-term contributions and relieves people of manual efforts, while generally retaining full control of when and what to contribute. 
Whenever a distributed dataflow is executed with a resource configuration, performance data as well as key characteristics are collected and pushed to our data distribution layer after job completion. 
The recorded performance data is ideally validated prior to publishing it to the P2P network.
Since our approach targets existing performance modeling solutions which mainly require and utilize workload monitoring already, the outlined workflow neither introduces any additional overhead nor formulates unrealistic requirements.

%% file: sections/4_evaluation.tex
\section{Preliminary Results}
\label{sec:evaluation}

In this section, we present an initial examination of our idea, both by means of prototypical implementation as well as simulation, and discuss our findings in detail.

\begin{figure}[h!]
    \centering
    \includegraphics[width=.55\columnwidth]{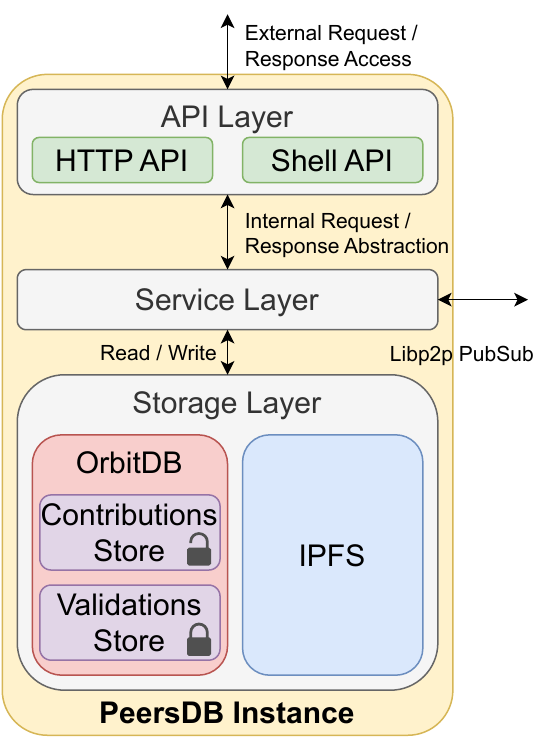}
    \caption{Architecture of the prototype implementation.
    The API layer facilitates access and usage of the developed service routines, which are in sync with other peers.
    Internally, the storage solution is based on IPFS, but regulates access to certain data and offers simplified, database-like means of interaction.}
    \label{fig:prototype}
\end{figure}

\subsection{Prototype}

We craft and test a first prototype that builds upon our previously introduced vision of a data distribution layer.

\textbf{Implementation.} For our prototype implementation, we employ the OrbitDB\footnote{\url{https://github.com/orbitdb/orbitdb}, Accessed: October 2023} project which uses IPFS in conjunction with Libp2p\footnote{\url{https://github.com/libp2p/go-libp2p}, Accessed: October 2023} to build a distributed database. 
It equips us with all the necessary features of a distributed database, including creating various types of data stores, automatic replication, and peer communication.
This enables additional functionalities like collaborative data validation.
The Go-based OrbitDB version is chosen due to the prevalence of Go in IPFS and its related packages.
With the employment of OrbitDB, we also enable the use of common database-like operations (e.g. \emph{put}, \emph{delete}) for different kinds of data stores (e.g. \emph{documents}, \emph{events}).
Our prototype, named \emph{PeersDB}\footnote{\url{https://github.com/PeersDB/peersdb}}, is accessed via custom APIs like HTTP and Shell, with extensibility options.
An illustration is provided with~\autoref{fig:prototype}.
APIs translate requests (e.g. \emph{get}, \emph{post}, \emph{query}) to an internal abstraction, suitable for the service component, forwarding them to the service Go-routine.
This routine manages recurring tasks like user requests, data storage, event handling, P2P communication for new peers, and collaborative validation coordination, thereby facilitating P2P-based communication and data exchange.
Accepting file contributions aligns with our system's purpose, adopting Filecoin's~\cite{PsarasD20} strategy by running an IPFS node for storing and referencing files through CIDs.
References are shared via OrbitDB among peers in the contributions store, fully replicated, granting access to training data without individual storage.
Said contributions store is realized by instantiation of an OrbitDB \emph{EventLogStore}, an append-only log with traversable history, which in turn uses the IPFS-Log internally.
The validations store, another service component, holds non-replicated validation details on user-selected training data.
It is based on the OrbitDB \emph{DocumentStore} and only used locally.
Both data stores use the same IPFS node as file storage, streamlining distribution.

\begin{table}[t]
\centering
\caption{Prototype: Hardware \& Software Specifications}
    \begin{tabular}[t]{rp{0.65\linewidth}}
        \toprule
        Resource&Details\\
        \midrule
        OS & Ubuntu 20.04\\
        CPU, vCores & Intel Cascade Lake, 2\\
        Memory, Network & 8 GB RAM, max. 4 Gbit/s outgoing\\
        Software & Golang 1.19, go-libp2p 0.27, kubo 0.19, Helm 3.12\\
        & go-orbit-db 1.21, Docker 20.10, Kubernetes 1.27\\
        \bottomrule
    \end{tabular}
\label{tbl:prototype_clusterspecs}
\end{table}

\textbf{Infrastructure Setup.} We evaluate our prototype in a Google Kubernetes Engine (GKE) cluster with six nodes of machine type \emph{e2-standard-2}, with one node each in the region \emph{asia-east2}, \emph{europe-west3}, \emph{us-west1}, \emph{south-america-east1}, \emph{me-west1}, and \emph{australia-southeast1}, to evaluate the system over long physical distances.
A single peer is then deployed to one of the nodes in the form of a Kubernetes Pod that holds the running PeersDB instance.
Targeted deployment to specific nodes and hence regions can be achieved by the utilization of node selectors.
To ease the execution of experiments, we design a Helm chart that allows for simplified parametrizations.
Further technical details can be found in~\autoref{tbl:prototype_clusterspecs}.

\textbf{Experiments.} We design and execute two basic experiments to assess the general feasibility of our system idea.
All necessary commands are executed against the API layer of our prototype to accurately measure the system performance.
In the first experiment, excessive amounts of data, precisely 11133 file uploads with an average compressed size of 9.06 Kb, are submitted into an already formed PeersDB cluster comprising 31 regular peers (distributed across regions) and one root peer (region asia-east2).
The focus here is on general replication metrics.
In the second experiment, 52 peers are added bit by bit to an already populated PeersDB cluster comprising initially of the root peer only.
In the beginning, they were added with a downtime of 1 minute between startups, which was reduced to 30 seconds after the first 12 peers. 
The chosen physical machine and therefore region were changed with every deployment to avoid resource contention between starting peers.
Here, we are most interested in the bootstrapping times of individual peers.
In both experiments, we make use of representative workload performance data of existing datasets\footnote{\url{https://github.com/dos-group/c3o-experiments}, Accessed: October 2023}\footnote{\url{https://github.com/oxhead/scout}, Accessed: October 2023} to make the experiments as natural as possible.
Note that we use a validation model in these experiments with a fairly constant response time (identity function), as it might otherwise influence the entire system's performance and hence corrupt measured metrics.

\begin{figure}
    \centering
    \includegraphics[width=\columnwidth]{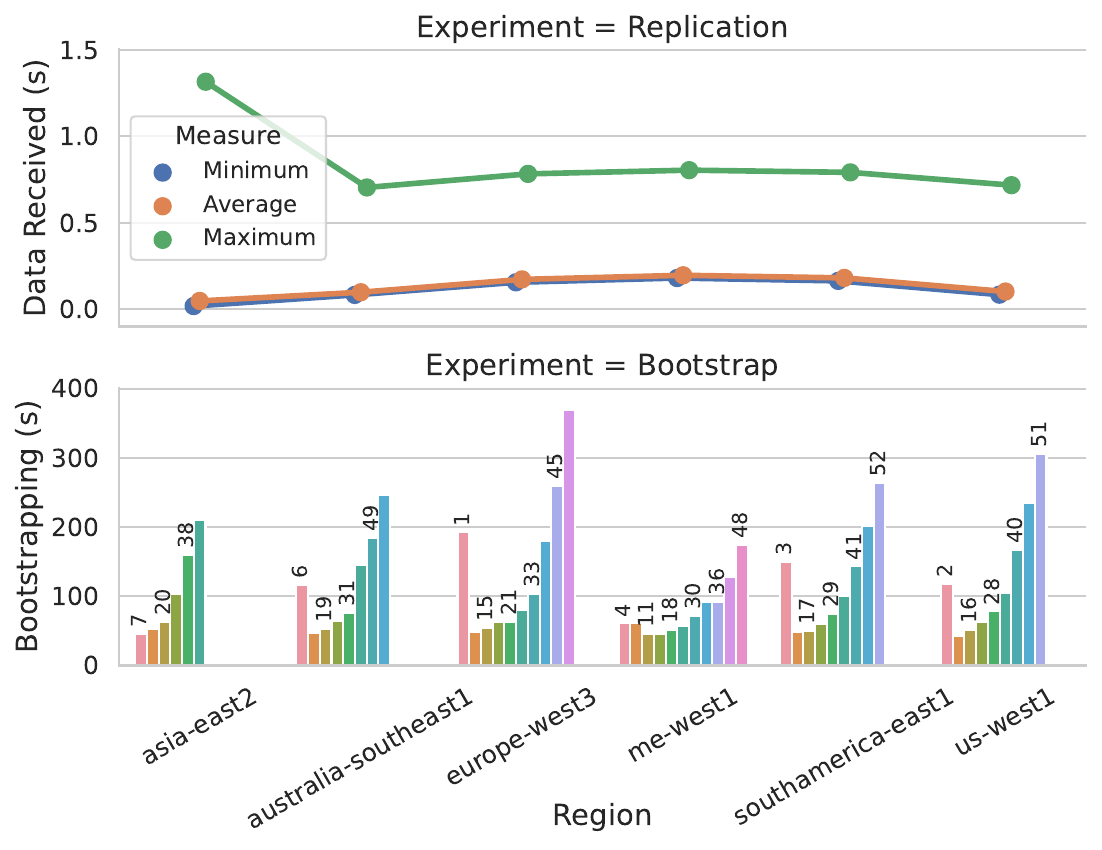}
    \caption{
    Results of our experiments with the prototype implementation.
    Small differences in data transmission times can be observed, as well as the bootstrapping time being conditioned on PeersDB cluster sizes. 
    }
    \label{fig:prototype_results}
\end{figure}

\textbf{Results.} We illustrate some of our results in~\autoref{fig:prototype_results}.
While the replication of 11133 files is a considerable load, the replication time of individual contributions across all nodes stays below one second in most instances, as can be seen in the top figure where the results of nodes are averaged per region.
It can also be observed that peers within one region demonstrate almost identical performance, with measurable differences only being worth mentioning when comparing regions against each other.
Here, smaller deviations can be noticed, although they do not seem to be due to physical distances.
The comparably high maximum replication time for region asia-east2 can be explained by the high CPU strain of the Kubernetes Pod containing the root peer due to, e.g., the bootstrapping process, which impacts also upon other Kubernetes Pods on the same physical machine.

The results of the second experiment are depicted in the bottom figure, where we annotated selected bars with numbers in order to show the size of the cluster at the time of joining.
We can make two observations from this plot: First, the overall size of the cluster impacts the bootstrapping time for every new peer to join -- this can be attributed to the communication overhead and other synchronization processes.
Another explanation is that peers on one node start at different times, resulting in varying availability of system resources.
Second, the bootstrapping time can be improved if there is already a peer geographically nearby where data can be pulled from.
While these results are to be expected, they showcase that our implementation follows an anticipated behavior. 

\subsection{Simulation}

For further testing of different performance data validation strategies in a controlled manner, we employ a simulation framework for distributed peer-to-peer systems.

\textbf{Implementation.} With the Testground\footnote{\url{https://docs.testground.ai}, Accessed: October 2023} platform for testing, benchmarking, and simulating distributed and peer-to-peer systems, we find a suitable tool for further investigating our system idea, especially since major building blocks of our idea such as IPFS have written test plans in Testground themselves, which allows for their adaptation as we see fit.
In particular, we adapt the \emph{bitswap-tuning} test plan\footnote{\url{https://github.com/ipfs/test-plans}, Accessed: October 2023} of IPFS and extend it in a way\footnote{\url{https://github.com/mcd01/test-plans}} that we can evaluate different data validation scenarios.

\begin{table}[t]
\centering
\caption{Simulation: Hardware \& Software Specifications}
    \begin{tabular}[t]{rp{0.65\linewidth}}
        \toprule
        Resource&Details\\
        \midrule
        OS & Ubuntu 20.04\\
        CPU, vCores & AMD EPYC 7282 2.8 GHz, 32\\
        Memory, Network & 128 GB RAM, 10 Gbit/s Ethernet NIC\\
        Software & Golang 1.20, Testground 0.6.0, Docker 20.10\\
        & go-libp2p 0.29, testground-sdk-go 0.3.0\\
        \bottomrule
    \end{tabular}
\label{tbl:simulation_clusterspecs}
\end{table}

\textbf{Infrastructure Setup.} We deploy our adapted test plan using Testground's docker runner, which ensures plans are built the same way with the same build environment even when built on a different machine or a different OS.
Since each container is lightweight by design, we can conduct experiments with a reasonable number of instances while simulating and observing an actually distributed peer-to-peer system.
Our test plan is executed on a private cluster node, with further technical details being provided in~\autoref{tbl:simulation_clusterspecs}.

\textbf{Experiments.} We execute the two original tests defined in the IPFS bitswap-tuning test plan, namely \emph{fuzz} (randomly disconnect and reconnect during transmission) and \emph{transfer} (transmission of differently sized files), together with manifold configurations of e.g. instance count, file sizes, latencies, jitter, or bandwidth limitations. 
In addition, we define multiple functions that mimic the scaling behavior (e.g. constant, linear, polynomial, exponential, logarithmic) of commonly used models for data validation in light of varying data amounts.
This way, we can investigate the scalability of our proposed validation component, and highlight potential limitations.

\textbf{Learnings.} The conducted experiments helped us better understand the intricacies of performance data validation in a decentralized manner, and manifested themselves in several important observations.
First, responses to validation requests in the network can and should be fast, which requires that validation processes run asynchronously in a background task.
A peer can thereby answer in a timely manner with its current validation information, which are updated in a non-blocking manner once the corresponding asynchronous validation task is finished.
This ensures that for our outlined workflows in~\autoref{sec:data_distribution_layer}, the communication overhead of the validation component remains manageable.
Second, as expected, different validation procedures exhibit different scaling behaviors, which might prolong the validation of single or multiple performance data points. 
For certain validation procedures, it might be worth considering batched performance data validation in order to accelerate the process.
Another tuning parameter is the number of responses from peers deemed sufficient in order to decide on a vote -- depending on the configuration, this might lead to a situation where peers can faster rely on the validation decisions of others, which prevents additional individual validation efforts.
Lastly, any candidate for a performance data validation strategy must guarantee to produce a deterministic outcome for a given performance data input, for it to work in a collaborative manner.

\subsection{Discussion}

With our experiments conducted with our prototype implementation as well as the simulation environment, we brought certain aspects of our approach to life and were able to assess them.
In this context, we observed that our prototype shows good performance with regard to fundamental functionalities such as replication, bootstrapping, and collaborative validation communication. 
Since our tackled use case does not assume high throughput loads and overly large clusters, we deem the performance acceptable, as we used realistic data in our experiments and scaled the prototype cluster up to around 50 peers while maintaining good results.
If the assumptions regarding the use case were to change in the future though, changes to our approach might become necessary.

We furthermore carved out some design considerations for our opportunistic decentralized data validation mechanism based on our simulation insights. 
In case stronger data validity guarantees were required, it would, however, be worth considering employing a blockchain to tackle this issue, at the cost of additional complexity, a less straightforward deployment, and reduced flexibility in the design of the data validation.

On another note, as with any P2P-based system, it is important to create incentives for participation.
Access to valuable performance data should be a good reason in the first place, but one might need to consider additional strategies in certain scenarios, e.g., to prevent short-lived participation.

%% file: sections/5_related_work.tex
\section{Related Work}
\label{sec:related_work}

This section discusses related methods regarding collaborative performance modeling and decentralized data exchange.

\subsection{Collaborative Workload Performance Modeling}

Since performance modeling of distributed dataflow jobs requires sufficient amounts of training data which is often not available, collaborative solutions represent an interesting approach to addressing this limitation.
In recent years, works have been proposed that either explicitly or by design qualify for application in collaborative scenarios.
For instance, multitask Bayesian Optimization or other ensemble methods are employed in~\cite{FekryCPR20,FekryCPRH20} to leverage performance data from executions of comparable workloads in order to find near-optimal resource configurations for distributed dataflows faster.
While the authors do not directly assume a collaborative scenario, these methods could be used in conjunction with a system where collaborators share performance data and other relevant information.
The same applies to methods originally designed for slightly different domains, for example~\cite{ZhangWCJT0Z021}, where a meta-learning approach is followed to optimize the resource utilization for cloud databases.
In our own previous work~\cite{ScheinertABWWT21,WillTSBK21,ScheinertWWTWK23}, we proposed methods for explicit collaboration, based on centralized storage solutions as well as modeling techniques that combine performance data from various execution contexts into a single model.
So far though, our work focused on model development and trace dataset analysis, leaving room for improvement, as an actual collaborative system has yet to be deployed and our past considerations evolved around central performance data repositories which introduce additional challenges.

In this work, we put emphasis on the actual data exchange which is an important foundation for collaborative performance modeling techniques.
Contrary to previous work, we investigate decentralized storage solutions, which are more in line with the oftentimes dynamic nature of collaboration.

\subsection{Decentralized Data Exchange}
In contrast to traditional data exchange architectures, which are typically in control of single authorities or entities, decentralized data exchange describes the distribution of data and control across a network of autonomous nodes. As a way to facilitate secure, transparent, and tamper-resistant data sharing among participants, technologies such as peer-to-peer networks, blockchains, and distributed ledger systems are often employed and can yield benefits for various domains.
The landscape of the aforementioned technologies has been extensively studied and mapped out in a plethora of survey papers that e.g. discuss blockchain-based decentralized storage~\cite{KhalidEAIHUN23, BenisiAJ20} in particular, or IPFS and P2P networks~\cite{DanielT22, NaikK20} in general. Additionally, numerous comparative studies addressed the performance of different decentralized data exchange approaches and shed light on the performance nuances of these systems by conducting extensive experiments~\cite{ShenLZW19, AscigilRKP0HKK19, HenningsenFR020, LajamH21}.

For instance, ChainSQL~\cite{MuzammalQN19} focuses on combining blockchain with traditional database approaches to securely store transaction logs, whilst leveraging the defined data structure of a database for more efficient queries. Other blockchain-based approaches aim to improve the system's performance by layering blockchains~\cite{AnielloBGLMS17}, to offer a platform for the trusted exchange of IoT data~\cite{8322729} or to enable the secure sharing of sensitive health data~\cite{AkkaouiHC20}.

However, although blockchain technology offers a promising perspective for decentralized trust-building and collaboration, it is not well-suited for the storage and exchange of possibly large performance data due to its inherent limitations in terms of size, complexity, and performance of the system.
For that reason, several approaches employ a combination of IPFS and blockchains in order to cope with these limitations. Sun et al.~\cite{DBLP:journals/ejwcn/SunHLWCW22} present an approach to securely store medical records in IPFS, while using a blockchain and smart contracts to enable fine-grained access control. Similarly, Randhir et al.~\cite{DBLP:conf/comsnets/KumarMT20} propose an off-chain storage of healthcare information in IPFS and leverage Proof-of-Identity and Proof-of-Work consensus methods offered by a blockchain.
Other works investigate the feasibility of using IPFS as a storage backend for sensory data collected in the agricultural domain and subsequently improving product tracking and food safety~\cite{hao2018safe} or to provide a decentralized application to safely store, manage, and share vehicle data~\cite{ye2021reliable}.

In our approach, we also utilize IPFS as a storage backend but neglect employing blockchain technology since it increases system complexity and imposes a substantial resource consumption overhead, which can hinder the efficiency of the overall solution.

%% file: sections/6_conclusion.tex
\section{Conclusion}
\label{sec:conclusion}

The primary goal of this work is to demonstrate the feasibility of a decentralized storage solution for collaborative resource optimization approaches of distributed dataflow applications.
To this end, we envision a system that relies on peer-to-peer communication and fosters performance data exchange among peers, such that individual performance modeling capabilities are improved while limitations of centralized storage solutions are overcome.
Towards this goal, we designed and implemented a decentralized collaboration approach based on an IPFS-backed database, and evaluated it as a prototype in experiments with real-world performance data, as well as using simulations for data validation strategies.
Our initial results showcase the scalability and reliability of our solution for the intended use case, supporting applicability in production.

In the future, we plan to further extend our prototype to improve its robustness, possibly by integration with blockchain technologies.
Moreover, we want to integrate it with existing performance modeling techniques and assess its performance in long-term experiments.